\documentclass[twocolumn,showpacs,preprintnumbers,amsmath,amssymb]{revtex4}

\newtheorem{theorem}{Theorem}
\newcommand{\Z}{{\mathbb Z}}

\newcommand{\C}{{\mathbb C}}

\begin{document}


\title{Dynamical Upper Bounds for the Fibonacci Hamiltonian}

\author{David Damanik}

\affiliation{Mathematics 253--37, California Institute of Technology, Pasadena, CA 91125,
USA.}

\email{damanik@caltech.edu}

\author{Serguei Tcheremchantsev}

\affiliation{UMR 6628--MAPMO, Universit\'{e} d'Orl\'{e}ans, B.P.~6759, F-45067
Orl\'{e}ans Cedex, France.}

\email{stcherem@univ-orleans.fr}

\date{\today}

\begin{abstract}
We consider transport exponents associated with the dynamics of a wavepacket in a
discrete one-dimensional quantum system and develop a general method for proving upper
bounds for these exponents in terms of the norms of transfer matrices at complex
energies. Using this method, we prove such upper bounds for the Fibonacci Hamiltonian.
Together with the known lower bounds, this shows that these exponents are strictly
between zero and one for sufficiently large coupling and the large coupling behavior
follows a law predicted by Abe and Hiramoto.
\end{abstract}


\maketitle

\section{\label{Sec1}Introduction}

The spreading of a quantum mechanical wavepacket has been the focus of intense research
both in the physics and mathematics communities. The free case and, more generally, the
case where the spectral measure of the initial state has an absolutely continuous
component are well understood. More recently, the case of point spectrum has received a
lot of attention and the question whether spectral localization leads to dynamical
localization has been answered quite satisfactorily \cite{djls}. As a consequence,
dynamical localization was shown in most systems for which spectral localization had been
established earlier. The most prominent exception to this rule is given by the random
dimer model \cite{dwp}, which displays spectral localization and dynamical delocalization
and, in fact, super-diffusive transport.

Between these two extreme cases there are a lot of models for which intermediate spectral
and dynamical behavior has been observed; for example, Bloch electrons in a magnetic
field \cite{pei} and one-dimensional quasicrystals \cite{kkt,sbgc}. The spreading of a
wavepacket in such systems is much less understood. There are a number of works based on
heuristics and numerics for the Harper and Fibonacci models devoted to this issue. As a
consequence it is expected that transport in the Harper model is almost diffusive and
transport in the Fibonacci model is anomalous in the sense that the transport exponents
take values other than zero (which would be the case in a dynamically localized system),
one-half (diffusive transport), and one (ballistic transport). In fact, these exponents
decrease as the coupling constant is increased and their behavior for large values of the
coupling constant $\lambda$ is expected to be of order $1/\log \lambda$ \cite{ah1,gkp}.

For the Harper model, there are no explicit rigorous bounds for the transport exponents.
Bellissard et al.\ show a lower bound in terms of the multifractal dimensions of the
density of states, \cite{bgs}, but they do not have lower bounds for these dimensions. On
the other hand, there are some explicit rigorous results for the Fibonacci model.
However, they are limited to bounding the transport exponents from below. The question of
whether the expected upper bounds, and hence anomalous transport, hold has been open from
a rigorous standpoint. In fact, there was no general method for proving upper bounds on
transport exponents, and finding such a method (that is applicable to models of interest)
is one of the most important problems in this field.

In any event, quantum dynamics for models with singular continuous spectra supported on
Cantor sets with critical eigenfunctions is a rich subject and has been studied by many
authors, with most papers focusing on the Harper model and the Fibonacci model or
generalizations thereof. A nice discussion of ``what determines the spreading of a
wavepacket,'' particularly in these intermediate cases, may be found in \cite{ketz}.

It is the purpose of this Letter to report on rigorous work concerning these issues. We
have developed a general approach for proving upper bounds on transport exponents in
terms of the norms of transfer matrices. These matrices are the standard tool, in one
dimension, to study spectral (and dynamical) properties of a given model. Essentially, it
is always the ultimate goal to reduce a problem at hand (such as studying dynamical
quantities) to properties of transfer matrices or, which is essentially equivalent,
solutions of the associated difference equation. The first rigorous results on lower
bounds for transport exponents could only be established after Jitomirskaya and Last had
found such a general correspondence \cite{JL}. See \cite{JL} for the first explicit lower
bounds for transport exponents in the Fibonacci model and \cite{DST} for the best bounds
known to this date. We will describe our general method for proving upper bounds on
transport exponents in Theorem~\ref{main1} below. Secondly, we have applied this method
to the Fibonacci model and proved upper bounds on transport exponents that are indeed of
order $1/\log \lambda$ in the large coupling regime. See Theorem~\ref{main2} for the
detailed statement. Together with the known lower bounds, this establishes the first
rigorous anomalous transport result.

The organizations is as follows: We recall the definitions of the transport exponents and
some of the known rigorous lower bounds for them in Section~\ref{Sec2}, describe our
general upper bound for these exponents in terms of transfer matrix norms in
Section~\ref{Sec3}, and discuss the application of this general result to the case of the
Fibonacci model in Section~\ref{Sec4}.

\section{\label{Sec2}Transport Exponents}

Consider a discrete one-dimensional Schr\"odinger operator,
\begin{equation}\label{oper}
[H \psi] (n)=\psi(n+1)+\psi(n-1) + V(n)\psi(n),
\end{equation}
on $\ell^2(\Z)$. A number of recent papers (e.g., \cite{DST,gcl,gs,JL,dkl,KKL,BGKT,GKT})
were devoted to proving lower bounds on the spreading of an initially localized
wavepacket, say $\psi = \delta_1$, under the dynamics governed by $H$, typically in
situations where the spectral measure of $\delta_1$ with respect to $H$ is purely
singular and sometimes even pure point.

A standard quantity that is considered to measure the spreading of the wavepacket is the
following: For $p > 0$, define $\langle |X|_{\delta_1}^p \rangle (T) = \sum_n |n|^p
a(n,T)$, where $a(n,T)=\frac{2}{T} \int_0^{\infty} e^{-2t/T} | \langle e^{-itH} \delta_1,
\delta_n \rangle |^2 \, dt$. Clearly, the faster $\langle |X|_{\delta_1}^p \rangle (T)$
grows, the faster $e^{-itH} \delta_1$ spreads out, at least averaged in time. One
typically wants to prove power-law bounds on $\langle |X|_{\delta_1}^p \rangle (T)$ and
hence it is natural to define the following quantity: For $p > 0$, define the
\textit{lower transport exponent} $\beta^-_{\delta_1}(p)$ by
$$
\beta^-_{\delta_1}(p)=\liminf_{T \to \infty} \frac{\log \langle |X|_{\delta_1}^p \rangle
(T) }{p \, \log T}
$$
and the \textit{upper transport exponent} $\beta^+_{\delta_1}(p)$ by
$$
\beta^+_{\delta_1}(p)=\limsup_{T \to \infty} \frac{\log \langle |X|_{\delta_1}^p \rangle
(T) }{p \, \log T}.
$$
Both functions $\beta^\pm_{\delta_1} (p)$ are non-decreasing. Numerical studies suggest
that for some models, they may exhibit non-trivial growth \cite{ketz,gm,p}. Such a
multiscaling phenomenon was termed \textit{quantum intermittency} by Guarneri and Mantica
\cite{gm}. For the two main models of interest, numerics show that there is non-trivial
growth for the Harper model, but no growth for the Fibonacci model \cite{ketz}.

Another way to describe the spreading of the wave-function, which turns out to capture
the limiting behavior of $\beta^\pm_{\delta_1} (p)$ for small and large values of $p$,
respectively, is in terms of probabilities. We define time-averaged \textit{outside
probabilities} by $P(N,T)=\sum_{|n|>N} a(n,T)$. Denote $S^-(\alpha) = - \liminf_{T \to
\infty} \frac{\log P(T^\alpha -2, T) }{\log T}$ and $S^+(\alpha) = - \limsup_{T \to
\infty} \frac{\log P(T^\alpha -2, T) }{\log T}$ for any $\alpha \in [0, +\infty]$. They
obey $0 \le S^+ (\alpha) \le S^- (\alpha) \le \infty$. These numbers control the power
decaying tails of the wavepacket. In particular, the following critical exponents are of
interest:
$$\alpha_l^\pm = \sup \{ \alpha \ge 0  : S^\pm (\alpha)=0 \}$$ and
$$\alpha_u^\pm = \sup \{ \alpha \ge 0  :  S^\pm (\alpha) < \infty \}.$$ We have that $0
\le \alpha_l^- \le \alpha_u^- \le 1$, $0 \le \alpha_l^+ \le \alpha_u^+ \le 1$, and also
that $\alpha_l^- \le \alpha_l^+$, $\alpha_u^- \le \alpha_u^+$.

One can interpret $\alpha_l^\pm$ as the (lower and upper) rates of propagation of the
essential part of the wavepacket, and $\alpha_u^\pm$ as the rates of propagation of the
fastest (polynomially small) part of the wavepacket. In particular, if
$\alpha>\alpha_u^+$, then $P(T^\alpha, T)$ goes to $0$ faster than any inverse power of
$T$. Since a ballistic upper bound holds in our case (for any potential $V$), Theorem 4.1
in \cite{GKT} yields $\lim_{p \to 0} \beta_{\delta_1}^\pm (p) = \alpha_l^\pm$ and
$\lim_{p \to \infty} \beta_{\delta_1}^\pm (p) = \alpha_u^\pm$. In particular, since
$\beta^\pm_{\delta_1} (p)$ are nondecreasing, we have that $\beta_{\delta_1}^\pm (p) \le
\alpha_u^\pm \quad \text{ for every } p > 0$.

When one wants to bound all these dynamical quantities for specific models, it is useful
to connect them to the qualitative behavior of the solutions of the difference equation
\begin{equation}\label{eve}
u(n+1) + u(n-1) + V(n)u(n) = z u(n)
\end{equation}
since there are effective methods for studying these solutions. Presently, the known
general results are limited to one-sided estimates of the transport exponents. Namely, a
number of approaches to lower bounds on $\beta^\pm_{\delta_1}(p)$ have been found in
recent years. The papers \cite{dkl,JL,KKL} derive such bounds in terms of the behavior of
solutions to \eqref{eve} with real energies $z$, with a link furnished by
Hausdorff-dimensional properties of spectral measures due to results of Guarneri, Combes,
and Last \cite{gcl}. In fact, what is proven in these papers (although not stated in this
form) are lower bounds on $\alpha_l^-$. Therefore, the lower bounds for
$\beta^-_{\delta_1} (p)$ obtained in this way are constant in $p$. There is also work by
Guarneri and Schulz-Baldes \cite{gs}, who bound $\alpha_l^+$ from below in terms of the
packing dimension of the spectral measure.

Further developments of these ideas by Guarneri and Schulz-Baldes, Barbaroux et al., and
Tcheremchantsev \cite{BGKT} allowed these authors to obtain better lower bounds for
$\beta_{\delta_1}^\pm (p)$ which are growing in $p$. These results elucidate the
phenomenon of quantum intermittency.

On the other hand, \cite{DST} develop a direct approach without an intermediate step.
These papers use power-law upper bounds on solutions corresponding to energies from a set
$S$ to derive lower bounds for $\beta^-_{\delta_1}(p)$. The set $S$ can even be very
small. One already gets non-trivial bounds when $S$ is not empty. If $S$ is not
negligible with respect to the spectral measure of $\delta_1$, the bounds are stronger,
but there are situations of interest (e.g., random polymer models \cite{dwp}), where the
spectral measure assigns zero weight to $S$.

\section{\label{Sec3}Transfer Matrices and Upper Bounds for Transport Exponents}

It should be stressed that there were no general rigorous methods for bounding
$\alpha_l^\pm, \alpha_u^\pm$, or $\beta^\pm_{\delta_1}(p)$ non-trivially from above. In
the present paper we propose a first general approach to proving upper bounds on
$\alpha_u^\pm$ (which in turn bound $\alpha_l^\pm$ and $\beta^\pm (p)$ for all $p>0$ from
above, as well). This approach relates the dynamical quantities introduced above to the
behavior of the solutions to the difference equation \eqref{eve} for complex energies
$z$. To state this result, let us recall the reformulation of \eqref{eve} in terms of
transfer matrices. These matrices are uniquely determined by the requirement that
$$
\left( \begin{array}{c} u(n+1) \\ u(n) \end{array} \right) = \Phi(n,z) \left(
\begin{array}{c} u(1) \\ u(0) \end{array} \right)
$$
for every solution $u$ of \eqref{eve}. Consequently,
$$
\Phi(n,z) = \begin{cases} T(n,z) \cdots T(1,z) & n \ge 1, \\ \mathrm{Id} & n = 0,
\\ [T(n+1,z)]^{-1} \cdots [T(0,z)]^{-1} & n \le -1, \end{cases}
$$
where
$$
T(m,z) = \left( \begin{array}{cr} z - V(m) & -1 \\ 1 & 0 \end{array} \right).
$$
We have the following result:

\begin{theorem}\label{main1}
Suppose $H$ is given by \eqref{oper}, where $V$ is a bounded real-valued function, and $K
\ge 4$ is such that $\sigma (H) \subseteq [-K+1,K-1]$. Suppose that, for some $C \in
(0,\infty)$ and $\alpha \in (0,1)$, we have
$$
\int_{-K}^K \left( \max_{3 \le n \le C T^\alpha} \left\| \Phi \left( n,E+ \tfrac{i}{T}
\right) \right\|^2 \right)^{-1} dE = O(T^{-m})
$$
and
$$
\int_{-K}^K \left( \max_{3 \le -n \le C T^\alpha} \left\| \Phi \left( n,E+ \tfrac{i}{T}
\right) \right\|^2 \right)^{-1} dE = O(T^{-m})
$$
for every $m \ge 1$. Then $\alpha_u^+ \le \alpha$. In particular, $\beta^+_{\delta_1} (p)
\le \alpha$ for every $p > 0$.
\end{theorem}

\noindent\textit{Remarks.} (a) If the conditions of the theorem are satisfied for some
sequence of times, $T_k \to \infty$, we get an upper bound for $\alpha_u^-$.\\[1mm]
(b) The proof of Theorem~\ref{main1} is based on the well-known formula $a(n,T) =
\frac{1}{T\pi} \int \left|\langle (H - E - \tfrac{i}{T})^{-1} \delta_1, \delta_n \rangle
\right|^2 \, dE$ and the fact that resolvent decay is closely related to lower bounds on
transfer matrix growth. Details will be given in \cite{DT2}.

\section{\label{Sec4}The Fibonacci Hamiltonian}

The Fibonacci Hamiltonian is an operator of the form \eqref{oper}, where the potential is
given by $V(n) = \lambda \chi_{[1- \alpha,1)}(n\alpha \mod 1)$ with $\alpha =
(\sqrt{5}-1)/2$. This potential belongs to the more general class of Sturmian potentials,
given by $V(n) = \lambda \chi_{[1- \alpha,1)}(n\alpha + \theta \mod 1)$ with general
irrational $\alpha \in (0,1)$ and arbitrary $\theta \in [0,1)$. These sequences provide
standard models for one-dimensional quasicrystals. (See \cite{sbgc} for the discovery of
quasicrystals.) Early  studies of the spectral properties of the Fibonacci model were
done by Kohmoto et al.\ and Ostlund et al.\ \cite{kkt}. It was suggested that the
spectrum is always purely singular continuous and of zero Lebesgue measure. This was
rigorously established by S\"ut\H{o} for the Fibonacci case \cite{su} and by Bellissard
et al.\ \cite{bist} and Damanik et al.\ \cite{dkl} in the general Sturmian case. Abe and
Hiramoto studied the transport exponents for the Fibonacci model numerically \cite{ah1};
see also Geisel et al.\ \cite{gkp}. They found that they are decreasing in $\lambda$ and
their work suggests that
\begin{equation}\label{fibdynexp}
\alpha_l^\pm , \, \alpha_u^\pm  \sim \frac{\mathrm{const}}{\log \lambda}
\end{equation}
as $\lambda \to \infty$.

The general approaches to lower bounds for the transport exponent described above have
all been applied to the Fibonacci Hamiltonian (and some Sturmian models). The best lower
bound for $\alpha_l^-$ was obtained by Killip et al.\ in \cite{KKL}. It reads $\alpha_l^-
\ge \frac{2 \kappa}{\zeta (\lambda) + \kappa+1/2}$, where $\kappa = \frac{\log
\frac{\sqrt{17} } {4} } {5 \log \left( \frac{\sqrt{5} + 1}{2} \right)} \approx 0.0126$
and $\zeta (\lambda)$, chosen so that one can prove a result like $\sum_{n = 1}^L \|\Phi
(n, E)\|^2 \le C L^{2 \zeta (\lambda) +1}$ for energies in the spectrum of $H$ (our
definition differs from that of \cite{KKL}), obeys $ \zeta (\lambda) = \frac{3 \log
\sqrt{5}}{\log \left( \frac{\sqrt{5} + 1}{2} \right)} \left( \log \lambda + O(1)
\right)$. This shows in particular that $\alpha_l^-$ admits a lower bound of the type
\eqref{fibdynexp}.

The best lower bound for $\alpha_u^-$ was found in \cite{DST}, where it was shown that
$\alpha_u^- \ge \frac{1}{\zeta (\lambda) + 1}$. In terms of the exponents
$\beta_{\delta_1}^-(p)$, the best known lower bounds are (see \cite{DST})
$$
\beta^-_{\delta_1}(p) \ge \begin{cases} \frac{p+2\kappa}{(p+1) (\zeta (\lambda) +
\kappa+1/2)}
& p \le 2 \zeta (\lambda)+1, \\
\frac{1}{\zeta (\lambda) + 1} & p>2 \zeta (\lambda) +1. \end{cases}
$$

We also want to mention work on upper bounds for the slow part of the wavepacket by
Killip et al.\ \cite{KKL}. More precisely, they showed that there exists a $\delta \in
(0,1)$ such that for $\lambda$ large enough, $P(C_2 T^{p (\lambda)}, T) \le 1 - \delta$.
Here, $p(\lambda) = \frac{6  \log  \frac{\sqrt{5} + 1}{2} } {\log \xi (\lambda)}$ and
\begin{equation}\label{xilambda}
\xi(\lambda) = \frac{\lambda - 4 + \sqrt{(\lambda - 4)^2 -12}}{2}.
\end{equation}
See \cite[Theorem~1.6.(i)]{KKL}. However, this result does not say anything about the
fast part of the wavepacket, and in particular, no statement for any of the transport
exponents can be derived.

With the help of Theorem~\ref{main1} we can prove upper bounds for $\alpha_u^+$ for the
Fibonacci model at sufficiently large coupling. These upper bounds show that
\eqref{fibdynexp} is indeed true.

The precise result is as follows:

\begin{theorem}\label{main2}
Consider the Fibonacci Hamiltonian and assume that $\lambda \ge 8$. Let $\alpha(\lambda)
= \frac{2 \log \frac{\sqrt{5}+1}{2}}{\log \xi(\lambda)}$, where $\xi(\lambda)$ is as in
\eqref{xilambda}. Then, $\alpha_u^+ \le \alpha (\lambda)$, and hence $\beta^+ (p) \le
\alpha (\lambda)$ for every $p > 0$.
\end{theorem}

\noindent\textit{Remarks.} (a) One can observe that $\alpha (\lambda) <
p(\lambda)$.\\[1mm]
(b) Note that $\xi(\lambda) = \lambda + O(1)$ as $\lambda \to \infty$. Moreover,
$\alpha(8) = \frac{2 \log \frac{\sqrt{5}+1}{2}}{\log 3} \approx 0.876$ and
$\alpha(\lambda)$ is a decreasing function of $\lambda$ for $\lambda \ge 8$. Thus, we
establish anomalous transport for the Fibonacci Hamiltonian with coupling $\lambda \ge 8$
and confirm the asymptotic dependence of the transport exponents $\alpha_u^\pm$ on the
coupling constant $\lambda$ that was predicted by Abe and Hiramoto. We emphasize again
that this is the first model for which anomalous transport can be shown
rigorously.\\[1mm]
(c) They key idea is to study the well-known trace map as a complex dynamical system and
prove upper bounds on the imaginary width of the (complex) canonical approximants of the
spectrum. Denote $x_n(z) = \mathrm{Tr} \, \Phi(F_n,z)$, where $F_n$ is the $n$-th
Fibonacci number and $z \in \C$, and $\sigma_n = \{ z \in \C : |x_n(z)| \le 2\}$. Then we
prove that $\sigma_n \subseteq \{ z \in \C : |\mathrm{Im} \, z| < C \xi(\lambda)^{-n/2}
\}$. Outside of the sets $\sigma_n \cup \sigma_{n+1}$, the traces grow
super-exponentially and this yields the lower bounds on the norms of transfer matrices
that we need. The claimed upper bound on $\alpha_u^+$ then follows from
Theorem~\ref{main1}. A detailed proof of Theorem~\ref{main2} may be found in \cite{DT2}.


\begin{thebibliography}{99}

\bibitem{djls} R.\ del Rio, S.\ Jitomirskaya, Y.\ Last, and B.\ Simon, Phys.\ Rev.\ Lett.\
\textbf{75}, 117 (1995); R.\ del Rio, S.\ Jitomirskaya, Y.\ Last, and B.\ Simon, J.\
Anal.\ Math.\ \textbf{69}, 153 (1996); S.\ Tcheremchantsev, Commun.\ Math.\ Phys.\
\textbf{221}, 27 (2001)

\bibitem{dwp} D.\ H.\ Dunlap, H.-L.\ Wu, and P.\ W.\ Phillips, Phys.\ Rev.\ Lett. {\bf 65}, 88
(1990); S.\ Jitomirskaya, H.\ Schulz-Baldes, and G.\ Stolz, Commun.\ Math.\ Phys.\ {\bf
233}, 27 (2003)

\bibitem{pei} R.\ Peierls, Z.\ Phys \textbf{80}, 763 (1933)

\bibitem{kkt} M.\ Kohmoto, L.\ P. Kadanoff, and C.\ Tang, Phys.\ Rev.\ Lett. \textbf{50}, 1870
(1983); S.\ Ostlund, R.\ Pandit, D.\ Rand, H.\ J.\ Schellnhuber, and E.\ D.\ Siggia,
Phys.\ Rev.\ Lett. \textbf{50}, 1873 (1983)

\bibitem{sbgc} D.\ Shechtman, I.\ Blech, D.\ Gratias, and J.\ V.\ Cahn, Phys.\ Rev.\
Lett.\ \textbf{53}, 1951 (1984)

\bibitem{ah1} S.\ Abe and H.\ Hiramoto, Phys.\ Rev.\ A \textbf{36}, 5349 (1987); H.\ Hiramoto and
S.\ Abe, J.\ Phys.\ Soc.\ Japan \textbf{57}, 230 (1988); H.\ Hiramoto and S.\ Abe, J.\
Phys.\ Soc.\ Japan \textbf{57}, 1365 (1988)

\bibitem{gkp} T.\ Geisel, R.\ Ketzmerick, and G.\ Petschel, in \textit{Quantum Chaos---Quantum Measurement}
(Copenhagen, 1991), 43, Kluwer, Dordrecht (1992)

\bibitem{bgs} J.\ Bellissard, I.\ Guarneri, and H.\ Schulz-Baldes, Commun.\ Math.\ Phys.\ \textbf{227}, 515 (2002)

\bibitem{ketz} R.\ Ketzmerick, K.\ Kruse, S.\ Kraut, and T.\ Geisel, Phys.\ Rev.\ Lett.\ \textbf{79},
1959 (1997)

\bibitem{JL} S.\ Jitomirskaya and Y.\ Last, Phys.\ Rev.\ Lett.\ \textbf{76}, 1765 (1999); S.\ Jitomirskaya
and Y.\ Last, Acta Math.\ {\bf 183}, 171 (1999); S.\ Jitomirskaya and Y.\ Last, Commun.\
Math.\ Phys.\ {\bf 211}, 643 (2000); for higher-dimensional analogues, see also A.\
Kiselev and Y.\ Last, Duke Math.\ J.\ \textbf{102}, 125 (2000)

\bibitem{DST} D.\ Damanik and S.\ Tcheremchantsev, Commun.\ Math.\
Phys.\ {\bf 236}, 513 (2003); D.\ Damanik, A.\ S\"ut\H{o}, and S.\ Tcheremchantsev, J.\
Funct.\ Anal.\ \textbf{216}, 362 (2004); S.\ Tcheremchantsev, Commun.\ Math.\ Phys.\
\textbf{253}, 221 (2005); D.\ Damanik and S.\ Tcheremchantsev, Preprint
(arXiv/math-ph/0407017)

\bibitem{gcl} I.\ Guarneri, Europhys.\ Lett.\ \textbf{10}, 95 (1989); J.\ M.\ Combes, in
\textit{Differential Equations with Applications to Mathematical Physics}, 59, Academic
Press, Boston (1993); Y.\ Last, J.\ Funct.\ Anal.\ \textbf{142}, 406 (1996)

\bibitem{gs} I.\ Guarneri and H.\ Schulz-Baldes, Math.\ Phys.\ Electron.\ J. \textbf{5}, Paper~1 (1999)


\bibitem{dkl} D.\ Damanik, R.\ Killip, and D.\ Lenz, Commun.\ Math.\ Phys.\ {\bf 212}, 191
(2000)

\bibitem{KKL} R.\ Killip, A.\ Kiselev, and Y.\ Last, Amer.\ J.\ Math.\ {\bf 125}, 1165 (2003)

\bibitem{BGKT} I. \ Guarneri and H. \ Schulz-Baldes, Lett.\ Math.\ Phys.\ \textbf{49}, 317 (1999);
J.-M.\ Barbaroux, F.\ Germinet, and S.\ Tcheremchantsev, Duke.\ Math.\ J.\ \textbf{110},
161 (2001); S.\ Tcheremchantsev, J.\ Funct.\ Anal.\ {\bf 197}, 247 (2003)

\bibitem{GKT} F.\ Germinet, A.\ Kiselev, and S.\ Tcheremchantsev, Ann.\ Inst.\ Fourier (Grenoble),
\textbf{54}, 787 (2004)


\bibitem{gm} I.\ Guarneri and G.\ Mantica, Phys.\ Rev.\ Lett.\ \textbf{73}, 3379 (1994)

\bibitem{p} F.\ Pi\'echon, Phys.\ Rev.\ Lett.\ \textbf{76}, 4372 (1996)

\bibitem{DT2} D.\ Damanik and S.\ Tcheremchantsev (to be published)

\bibitem{su} A.\ S\"ut\H{o}, Commun.\ Math.\ Phys.\ {\bf 111}, 409 (1987); A.\ S\"ut\H{o}, J.\
Statist.\ Phys.\ \textbf{56}, 525 (1989)

\bibitem{bist} J.\ Bellissard, B.\ Iochum, E.\ Scoppola, and D.\ Testard,
Commun.\ Math.\ Phys.\ {\bf 125}, 527 (1989)

\end{thebibliography}
\end{document}